\documentclass[12pt]{article}

\usepackage{graphicx}
\usepackage{endfloat}
\usepackage{amssymb}
\usepackage[english]{babel}
\usepackage{amsmath,amsthm}
\usepackage{amsfonts}
\usepackage{mathtools}
\usepackage{mathrsfs}
\usepackage{algorithm}
\usepackage[noend]{algpseudocode}
\usepackage{enumerate}
\usepackage{color}

\usepackage[default]{jasa_harvard}    
\usepackage{JABES_manu}

\usepackage[mathlines]{lineno}

\makeatletter
\def\BState{\State\hskip-\ALG@thistlm}
\makeatother

\begin{document}

\bibliographystyle{ECA_jasa}

\title{Dynamic Models of Animal Movement with Spatial Point Process Interactions}
\author{James C Russell, Ephraim M Hanks, and Murali Haran}
\maketitle

\begin{abstract}
 When analyzing animal movement, it is important to account for interactions between individuals. However, statistical models for incorporating interaction behavior in movement models are limited. We propose an approach that models dependent movement by augmenting a dynamic marginal movement model with a spatial point process interaction function within a weighted distribution framework. The approach is flexible, as marginal movement behavior and interaction behavior can be modeled independently. Inference for model parameters is complicated by intractable normalizing constants. We develop a double Metropolis-Hastings algorithm to perform Bayesian inference. We illustrate our approach through the analysis of movement tracks of guppies (\textit{Poecilia reticulata}).
\end{abstract}

\noindent\textsc{Keywords}: {auxiliary variable MCMC algorithm, collective motion, biased correlated random walk, group navigation, \textit{Poecilia reticulata}, state-space model}

\section{Introduction}\label{intro}

    Movement models are important for studying animal behavior as they can reveal how animals use space and interact with the environment. Information on the movement patterns of animal species can play an important role in conservation, particularly for migratory species \cite**{durban2012antarctic}. Many methods exist for modeling individual animal movement, including models that account for changing behaviors at different locations and times by utilizing Markovian switching models (e.g. \citeasnoun{Harris_2013}; \citeasnoun{Blackwell_1997}) and models that account for the animal's preferences for covariates measured throughout the territory (e.g. \citeasnoun**{Hooten_2013}; \citeasnoun**{Johnson_2013}).

    Interactions between animals can give insight into the structures of animal societies \cite**{mersch2013tracking}. Animal species often exhibit herd or school behavior, and even those that do not form groups have movement that depends on the behavior of other individuals. \citeasnoun**{Langrock} incorporate dependence by assuming the animals in a herd move around a central point, such as a designated group leader or a latent central location. \citeasnoun{codling2014} propose a model that combines individual navigational behavior with the tendency to copy the behavior of other nearby individuals by taking a weighted average of the two behavioral mechanisms. This enables information sharing among neighbors. \citeasnoun**{perna2014} consider a model that encourages individuals to have a preferential structure. For example, an individual might tend to stay directly behind another, thus creating a leader-follower relationship. \citeasnoun{sumpter2006} gives a broad overview of animal movement, including computer simulation models which utilize self propelled particle (SPP) systems with specific movement rules to account for interaction.

    We propose a model that describes continuous-time dynamics of animal movement \cite**{Johnson2008} while simultaneously allowing for current-location based interactions by modeling animal locations as a spatial interacting point process \cite{moller2004statistical}. Point process models allow interaction between animal locations such as clustering, regularity, or repulsion, through the use of interaction functions. This provides a paradigm for modelling different types of interactions between animals including collision avoidance, herding behavior, animals that break off into multiple smaller groups, and animals that interact with each other without moving in herds or schools. Our model uses a weighted distribution approach to incorporate several features, including
    \begin{enumerate}[i.]
    \itemsep0em
    \item directional persistence through a continuous-time biased correlated random walk,
    \item inter-animal behavior modeled using spatial point process interaction functions,
    \item observation error in animal locations.
    \end{enumerate}
    Other models exist which incorporate one or more of these features; we propose a flexible framework for all three.
\begin{figure}
\centering
\caption{Group Movement Paths}
\label{fig:SimulatedPaths}
\centering
\includegraphics[width=110mm]{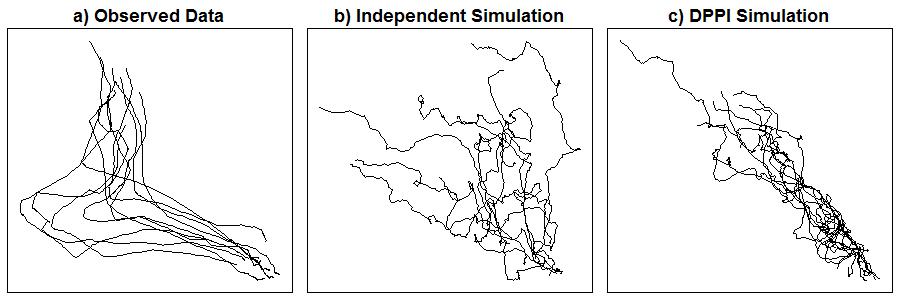}\\
\begin{flushleft}
a)Plotted paths of a shoal of 10 guppies from \citeasnoun**{Bode2012}. \\
b)Plotted paths of a simulated realization from the CTCRW model without interactions. \\
c)Plotted paths of a simulated realization from the DPPI model with the attraction-repulsion point process interaction function.
\end{flushleft}
\end{figure}

   To illustrate our approach we analyze the guppy (\textit{Poecilia reticulata}) movement data of \citeasnoun**{Bode2012} in which ten guppies are released in the lower right section of a fish tank, and are attracted to the top left by shelter in the form of shade and rocks. A realization of this experiment is shown in Figure \ref{fig:SimulatedPaths}(a) where the interaction between guppies is evident, as the guppies remain together in a shoal. To illustrate the need for statistical models incorporating between-animal dependence, Figure \ref{fig:SimulatedPaths}(b) shows a simulation from an independent movement model, as described in Section 2.2. In the simulation, the guppies tend to drift apart, so the model does not replicate the shoaling behavior. In Figure \ref{fig:SimulatedPaths}(c) we show a simulated realization from our proposed dynamic point process interaction (DPPI) model, described in Section 2.4. Each guppy's marginal movement is modeled as a continuous-time biased correlated random walk which results in smooth paths similar to the observed guppy paths. Group movement is modeled using the attraction-repulsion interaction function of \citeasnoun**{Goldstein}. The simulated guppies in Figure \ref{fig:SimulatedPaths}(c) stay together in a group, similar to the observed guppies in  Figure \ref{fig:SimulatedPaths}(a).

    The rest of this paper is organized as follows. In Section 2, we introduce the general modeling framework, and give several examples of point process interaction functions useful for modeling group animal movement. In Section 3, we propose a Markov Chain Monte Carlo algorithm to sample from the posterior distributions of model parameters. We describe a double Metropolis-Hastings algorithm for inference complicated by the intractable normalizing function that arises from our point process interaction approach to modeling group movement. In Section 4, we examine the performance of our approach by utilizing several simulated movement paths. Finally, in Section 5, we use our approach to analyze the guppy movement paths of \citeasnoun{Bode2012}.

\section{Modeling Movement Dynamics with Interactions}\label{sec1}

    In this section, we describe our proposed approach, starting with a continuous-time stochastic model for the dynamics of individual guppy movement. Next, we aggregate the individual model to incorporate multiple individuals and describe our point process approach to modeling interactions. Finally, we compare our approach to existing methods.

    Let the unobserved states, consisting of the true locations and instantaneous velocities, of individuals $(1,...,K)$ at a given time $t_i$ be denoted by $\boldsymbol{A}_{t_i} = (\boldsymbol{\alpha}^{(1)}_{t_i},\boldsymbol{\alpha}^{(2)}_{t_i},...,\boldsymbol{\alpha}^{(K)}_{t_i})^T$, and let $\boldsymbol{\Theta}$ denote our vector of parameters. We can write an aggregate group movement model by assuming independence and multiplying the marginal densities
    \begin{linenomath*}
    \begin{align*}
    f(\boldsymbol{A}_{t_i}|\boldsymbol{A}_{t_{i-1}},\boldsymbol{\Theta}) = \prod_{k=1}^K f(\boldsymbol{\alpha}^{(k)}_{t_i}|\boldsymbol{\alpha}^{(k)}_{t_{i-1}},\boldsymbol{\Theta})
    \end{align*}
    \end{linenomath*}
    where $f(\boldsymbol{\alpha}^{(k)}_{t_i}|\boldsymbol{\alpha}^{(k)}_{t_{i-1}},\boldsymbol{\Theta})$ represents a marginal movement model. That is, the $k^{th}$ individual's state at time $t_i$, $\boldsymbol{\alpha}^{(k)}_{t_i}$, is modeled conditional on that individual's state at time $t_{i-1}$, $\boldsymbol{\alpha}^{(k)}_{t_{i-1}}$, and the k individuals move independently of each other. To model movement interactions, we multiply the marginal model by an interaction function, which is a function of the pairwise distance between observations at time $t_i$, yielding a joint distribution
    \begin{linenomath*}
    \begin{align*} 
    f(\boldsymbol{A}_{t_i}|\boldsymbol{A}_{t_{i-1}},\boldsymbol{\Theta}) = \frac{\prod_{k=1}^K f(\boldsymbol{\alpha}^{(k)}_{t_i}|\boldsymbol{\alpha}^{(k)}_{t_{i-1}},\boldsymbol{\Theta}) \prod_{j<k}\psi_{jk}(\boldsymbol{\alpha}_{t_{i}}^{(j)}, \boldsymbol{\alpha}_{t_{i}}^{(k)};\boldsymbol{\Theta})}{c(\boldsymbol{\Theta})}
    \end{align*}
    \end{linenomath*}
    where $\psi_{jk}(\boldsymbol{\alpha}_{t_{i}}^{(j)}, \boldsymbol{\alpha}_{t_{i}}^{(k)};\boldsymbol{\Theta})$ is the interaction function, which we take from methods in point process literature. The resulting model is similar to the weighted distribution approach to modeling animal movement. \citeasnoun**{johnson2008b} and \citeasnoun**{lele2006} utilize this approach to model a resource selection function for animal telemetry data which accounts for animals preferentially selecting certain habitats. In our method, the animal's proximity to neighbors, rather than habitat resource covariates, are driving movement behavior. Note that $c(\boldsymbol{\Theta})$ is an intractable normalizing function of $\boldsymbol{\Theta}$. This complicates posterior evaluation as we will see later.

\subsection{Marginal Movement Model}

 To develop a group movement model with interactions, we start with an existing movement model for an individual, the continuous time biased correlated random walk model (CTCRW) from \citeasnoun**{Johnson2008}. The CTCRW model specifies an Ornstein-Uhlenbeck model for velocity, resulting in movement paths that show directional persistence, similar to that of the observed guppy movement paths in Figure 1(a). While not important for the guppy data, an additional advantage of the CTCRW model is that it allows for observations at non-uniform time points. The CTCRW model is flexible, and can easily be adjusted to account for complexities in a given data set. For example, \citeasnoun{durban2012antarctic} use the CTCRW model to estimate the displacement velocities of killer whalers; \citeasnoun**{citta2013dive} use an adjusted version of the CTCRW model to analyze haul out behavior of Eastern Chukchi beluga whales and \citeasnoun**{kuhn2014evidence} use the CTCRW model to estimate locations of northern fur seals along foraging tracks.

 Let $x(t)$ and $y(t)$ be the observed location coordinates of the animal at time $t$,  $\mu^{(x)}(t)$ and $\mu^{(y)}(t)$ be the true unobserved $x$ and $y$ locations of the animal at time $t$, and $v^{(x)}(t)$ and $v^{(y)}(t)$ the instantaneous $x$ and $y$ directional velocities of the animal at time $t$. Let $\boldsymbol{s}(t)$ be the observed location and $\boldsymbol{\alpha}_t$ the unobserved state at time $t$, with
 \begin{linenomath*}
 \begin{align}\label{eq:2.1}
            \boldsymbol{s_t} =  \left( \begin{array} {c} x(t) \\ y(t) \end{array} \right),  &&
            \boldsymbol{\alpha}_t =  \left( \begin{array} {c} \mu^{(x)}(t) \\ v^{(x)}(t) \\ \mu^{(y)}(t) \\ v^{(y)}(t)  \end{array} \right).
 \end{align}
 \end{linenomath*}

  We assume that $t \in \mathbb{R}^+$, and the locations $(x(t), y(t))$ belong to $\mathbb{R}^2$. The x and y elements are assumed to be independent, as a positive correlation between x and y velocities, for example, would indicate movement in a northeast or southwest direction.

 To model directional persistence in movement, $v^{(x)}(t)$ and $v^{(y)}(t)$ are assumed to follow independent continuous-time Ornstein-Uhlenbeck processes. We first present the CTCRW model for one-dimensional movement, focusing on the $x$ coordinate of Equation \eqref{eq:2.1}. Our development follows that of \citeasnoun{Johnson2008}.

 Given a change in time $\Delta$, the $x$-directional velocity is given by
 \begin{linenomath*}
 \begin{align}\label{eq:vel}
 v^{(x)}(t+\Delta) = \gamma_1 + e^{-\beta\Delta}[v^{(x)}(t)-\gamma_1] + \xi_1(\Delta),
 \end{align}
 \end{linenomath*}
where $\xi_1(\Delta)$ is a normal random variable with mean 0 and variance $\sigma^2 [1-\exp(-2\beta\Delta)]/2\beta$, $\sigma^2$ represents the variability in the random velocity, $\gamma_1$ describes the directional drift (mean velocity) in the $x$ direction, and $\beta$ controls the autocorrelation in velocity. Equation \eqref{eq:vel} reveals that the updated velocity at time $t + \Delta$ ($ v^{(x)}(t+\Delta)$) is equal to a weighted average of the mean drift ($\gamma_1$), and the velocity at time $t$ ($v^{(x)}(t)$) plus a random term with mean $0$. Using this parametrization, small values of $\beta$ imply a higher tendency to continue traveling with the same velocity over time. The location $\mu^{(x)}(t+\Delta)$ is obtained by integrating velocity over time
\begin{linenomath*}
\begin{align*}
\mu^{(x)}(t+\Delta) = \mu^{(x)}(t) + \int_{t}^{t+\Delta} v^{(x)}(u) du.
\end{align*}
\end{linenomath*}

  Assuming we have $N$ observations at times $(t_1,..., t_N)$ , discretization of the continuous time model yields the distributions for the unobserved states,
  \begin{linenomath*}
  \begin{align}\label{eq:2.2}
            \left( \begin{array} {c} \mu^{(x)}_{t_i} \\ v^{(x)}_{t_i} \end{array} \right) &\sim
            N\left(\boldsymbol{T_1}(\beta, \Delta_i) \left( \begin{array} {c} \mu^{(x)}_{t_{i-1}} \\ v^{(x)}_{t_{i-1}} \end{array} \right) + \boldsymbol{d_1}(\gamma_1, \beta ,\Delta_i ) ,\sigma^2 \boldsymbol{V_1}(\beta, \Delta_i)\right), \hfill i=1,...,N,
 \end{align}
 \end{linenomath*}
 where $\Delta_i$ is the time change between observations $i-1$ and $i$, $\boldsymbol{T_1}(\beta, \Delta_i)$ accounts for the directional persistence,
 \begin{linenomath*}
 \begin{align*}
            \boldsymbol{T_1}(\beta, \Delta_i) = \left( \begin{array} {cc} 1 & \frac{1-e^{-\beta\Delta_i}}{\beta} \\
            0 & e^{-\beta\Delta_i} \end{array} \right),
 \end{align*}
 \end{linenomath*}
 $\boldsymbol{d_1}(\gamma_1, \beta ,\Delta_i )$ models directional drift,
 \begin{linenomath*}
 \begin{align*}
            \boldsymbol{d_1}( \gamma_1, \beta ,\Delta_i ) =
            \gamma_1 \left( \begin{array} {c} \Delta_i - \frac{1-e^{-\beta\Delta_i}}{\beta} \\
            1 - e^{-\beta\Delta_i} \end{array} \right),
 \end{align*}
 \end{linenomath*}
 and the variance matrix of Equation \eqref{eq:2.2} is given by
 \begin{linenomath*}
 \begin{align*}
            \boldsymbol{V_1}(\beta, \Delta_i) &= \left( \begin{array} {cc} v_1(\beta, \Delta_i) & v_3(\beta, \Delta_i)
            \\ v_3(\beta, \Delta_i) & v_2(\beta, \Delta_i) \end{array} \right),
 \end{align*}
 \end{linenomath*}
 with
 \begin{linenomath*}
 \begin{align*}
            v_1(\beta, \Delta_i)&=\frac{\Delta_i - \frac{2}{\beta}(1-e^{-\beta\Delta_i}) + \frac{1}{2\beta}(1-e^{-2\beta\Delta_i})}{\beta^2}, \\
            v_2(\beta, \Delta_i)&= \frac{1-e^{-2\beta\Delta_i}}{2\beta},\\
            v_3(\beta, \Delta_i)&= \frac{1 - 2e^{-\beta\Delta_i} + e^{-2\beta\Delta_i}}{2\beta^2}.
 \end{align*}
 \end{linenomath*}
 Finally, the observed position ($s^{(x)}_{t_i}$) of the animal is modeled as a Gaussian random variable centered at the true location ($\mu^{(x)}_{t_i}$)
 \begin{linenomath*}
 \begin{align*}
            s^{(x)}_{t_i} &\sim N(\mu^{(x)}_{t_i},\sigma_E^2),
 \end{align*}
 \end{linenomath*}
 where $\sigma_E^2$ represents the observation error variance. To aggregate the x and y dimensional distributions into a 2-dimensional model, as given in Equation \eqref{eq:2.1}, the covariance terms between all x and y elements are set to 0. This yields the marginal model for the individual, with parameters $(\beta, \gamma_1, \gamma_2, \sigma^2, \sigma_E^2)$ and distributions
 \begin{linenomath*}
 \begin{align}\label{eq:2.4}
        \boldsymbol{s}_{t_i} &\sim N(\boldsymbol{Z}\boldsymbol{\alpha}_{t_i},\sigma_E^2 \boldsymbol{I_2})\\ \label{eq:2.4b}
        \boldsymbol{\alpha}_{t_i} &\sim N(\boldsymbol{T}(\beta, \Delta_i)\boldsymbol{\alpha}_{t_{i-1}} +
        \boldsymbol{d}(\gamma_1, \gamma_2, \beta ,\Delta_i ) ,\sigma^2 \boldsymbol{V}(\beta, \Delta_i)).
 \end{align}
 \end{linenomath*}
 where $\boldsymbol{T} = \boldsymbol{I_2} \otimes \boldsymbol{T_1}(\beta, \Delta_i)$, $\boldsymbol{d} = [  \boldsymbol{d_1}( \gamma_1, \beta ,\Delta_i )',  \boldsymbol{d_1}( \gamma_2, \beta ,\Delta_i )' ]'$, $\boldsymbol{V} = \boldsymbol{I_2} \otimes \boldsymbol{V_1}(\beta, \Delta_i)$, and
 \begin{linenomath*}
 \begin{align*}
           \boldsymbol{Z} &= \left( \begin{array} {cccc} 1 & 0 & 0 & 0
            \\ 0 & 0 & 1 & 0 \end{array} \right),
 \end{align*}
 \end{linenomath*}
For details about the derivation of the model and examples using this model see \citeasnoun{Johnson2008}.

\subsection{Independent Group Movement Model}

Assuming independent movement between individuals, this model can be easily extended to a group setting. For the remainder of the article we assume that the movement parameters $(\beta, \gamma_1, \gamma_2, \sigma^2, \sigma_E^2)$ are shared by all individuals.

Assume that we observe $K \geq 1$ animals where every individual is observed at each time point $(t_1, t_2, ..., t_N)$. The observed locations are denoted by $\boldsymbol{S}_{t_i}= (\boldsymbol{s}^{(1)}_{t_i},\boldsymbol{s}^{(2)}_{t_i},...,\boldsymbol{s}^{(K)}_{t_i})^T$ for $t_i \in {t_1, t_2, ..., t_N}$ and the unobserved states are denoted $\boldsymbol{A_{t_i}} = (\boldsymbol{\alpha}^{(1)}_{t_i},\boldsymbol{\alpha}^{(2)}_{t_i},...,\boldsymbol{\alpha}^{(K)}_{t_i})^T$. The joint distribution for the unobserved states may be expressed as
\begin{linenomath*}
 \begin{align}\label{eq:2.6}
     g \left(\boldsymbol{A_{t_{1:N}}} |\beta, \gamma_1, \gamma_2, \sigma^2 \right) = \prod_{i=1}^{N} \prod_{k=1}^{K} f(\boldsymbol{\alpha}^{(k)}_{t_i}|\boldsymbol{\alpha}^{(k)}_{t_{i-1}}, \beta, \gamma_1, \gamma_2, \sigma^2),
\end{align}
\end{linenomath*}
where $f(\boldsymbol{\alpha}^{(k)}_{t_i}|\boldsymbol{\alpha}^{(k)}_{t_{i-1}}, \beta, \gamma_1, \gamma_2, \sigma^2)$ is the density of a normal random variable for the unobserved state for individual $k$ at time $t_i$, as defined in Equation \eqref{eq:2.4b}. The joint distribution for the observed locations conditional on the unobserved states is therefore
\begin{linenomath*}
 \begin{align}\label{eq:2.6b}
      h\left(\boldsymbol{S_{t_{1:N}}} |\boldsymbol{A_{t_{1:N}}}, \sigma^2_E \right) = \prod_{i=1}^{N} \prod_{k=1}^{K} f(\boldsymbol{s}^{(k)}_{t_i}|\boldsymbol{\alpha}^{(k)}_{t_{i}}, \sigma^2_E),
\end{align}
\end{linenomath*}
where $f(\boldsymbol{s}^{(k)}_{t_i}|\boldsymbol{\alpha}^{(k)}_{t_i}, \sigma_E^2)$ is the density of a normal random variable for the observation error for individual $k$ at time $t_i$, as defined in Equation \eqref{eq:2.4},

\subsection{Dynamic Point Process Interaction (DPPI) Model}

If we assume independence between individuals, once two animals start to drift apart, there is no mechanism to draw the animals back towards each other. To model schooling or herd behavior, we propose an approach motivated by spatial point process models. Consider Equation \eqref{eq:2.6}, which gives the distribution of the unobserved states of a set of animals at the current time point conditional on the locations at the previous time point. To simplify notation, let $\boldsymbol{\Theta_1} = (\beta, \gamma_1, \gamma_2, \sigma^2, \sigma_E^2)$ describe the parameters for the marginal movement model, and let $\boldsymbol{\Theta_2}$ describe the parameters for a spatial point process interaction function $\psi(\cdot)$. For each pair of locations at the current time point, we multiply the density by a point process interaction function $\psi_{jk}\left(\delta \left(\boldsymbol{\alpha}_{t_{i}}^{(j)}, \boldsymbol{\alpha}_{t_{i}}^{(k)}\right);\boldsymbol{\Theta_2}\right)$ which depends only on the pairwise Euclidean distance between the current locations, which we define to be $\delta \left(\boldsymbol{\alpha^{(j)}},\boldsymbol{\alpha^{(k)}}\right) = \sqrt{(\mu_x^{(j)}-\mu_x^{(k)})^2 + (\mu_y^{(j)}-\mu_y^{(k)})^2}$, and parameter $\boldsymbol{\Theta_2}$. Note that this is not a function of the unobserved velocities. Hence we multiply Equation \eqref{eq:2.6} by the product of our interaction functions
\begin{linenomath*}
\begin{align}\label{eq:psi}
\psi(\boldsymbol{A_{t_{1:N}}};\boldsymbol{\Theta_2}) =\prod_{i-1}^N \prod_{k=2}^{K} \prod_{j<k}\psi_{jk}\left(\delta \left(\boldsymbol{\alpha}_{t_{i}}^{(j)}, \boldsymbol{\alpha}_{t_{i}}^{(k)}\right);\boldsymbol{\Theta_2}\right)
\end{align}
\end{linenomath*}
which takes values in $\mathbb{R}^+$. For two animals $i$ and $j$, if the value of $\psi_{jk} ( \cdot )$ is small, this discourages animals from moving to these locations at the same time, similar to a weighted distribution approach for resource selection \cite**{johnson2008b}. The ordering of the individuals does not impact the results.

The resulting model has joint density given by:
\begin{linenomath*}
 \begin{align}\label{eq:density}
             \frac{ h\left(\boldsymbol{S_{t_{1:N}}} |\boldsymbol{A_{t_{1:N}}}, \sigma^2_E \right) g\left(\boldsymbol{A_{t_{1:N}}} |\beta, \gamma_1, \gamma_2, \sigma^2 \right) \psi( \boldsymbol{A_{t_{1:N}}} |\boldsymbol{\Theta_2})}{c(\boldsymbol{\Theta_1}, \boldsymbol{\Theta_2})},
 \end{align}
 \end{linenomath*}
where $h\left(\boldsymbol{S_{t_{1:N}}} |\boldsymbol{A_{t_{1:N}}}, \sigma^2_E \right)$ represents the density of the observed locations conditional on the unobserved states from Equation \eqref{eq:2.6b}, $g\left(\boldsymbol{A_{t_{1:N}}} |\beta, \gamma_1, \gamma_2, \sigma^2 \right)$ represents the density of the unobserved states from Equation \eqref{eq:2.6}, $\psi( \boldsymbol{A_{t_{1:N}}} |\boldsymbol{\Theta_2})$ represents the interaction function from Equation \eqref{eq:psi} and $c(\boldsymbol{\Theta_1}, \boldsymbol{\Theta_2})$ is the normalizing function required to ensure that the density integrates to 1 and is given by the multidimensional integral over the unobserved states:
\begin{linenomath*}
  \begin{align*}
             c(\boldsymbol{\Theta_1}, \boldsymbol{\Theta_2}) = \int h\left(\boldsymbol{S_{t_{1:N}}} |\boldsymbol{A_{t_{1:N}}}, \sigma^2_E \right) g\left(\boldsymbol{A_{t_{1:N}}} |\beta, \gamma_1, \gamma_2, \sigma^2 \right) \psi( \boldsymbol{A_{t_{1:N}}} |\boldsymbol{\Theta_2})d\boldsymbol{A_{t_{1:N}}}
 \end{align*}
\end{linenomath*}
The point process interaction function $\psi( \boldsymbol{A_{t_{1:N}}} |\boldsymbol{\Theta_2})$ should be selected based on the assumed interaction behavior of the animals being studied.

Herding or schooling behavior can be generated when individuals repel each other at small distances to avoid collisions, attract each other at mid range distances, and behave independently when they are a large distance apart. An interaction function that captures this behavior is the attraction-repulsion interaction function found in \citeasnoun{Goldstein}. This interaction function is given by:
\begin{linenomath*}
\begin{align*}
\psi(\boldsymbol{A_{t_{1:N}}} , \theta_1, \theta_2, \theta_3, R) = \prod_{i=1}^N \prod_{k=2}^{K} \prod_{j<k} \psi \left(\delta\left(\boldsymbol{\alpha}_{t_i}^{(j)}, \boldsymbol{\alpha}_{t_i}^{(k)}\right) ;\theta_1, \theta_2, \theta_3, R \right),
\end{align*}
\end{linenomath*}
with
\begin{linenomath*}
\begin{align}\label{eq:2.7}
\psi(r ;\theta_1, \theta_2, \theta_3, R) = \begin{cases} 0 &\mbox{if } 0 \leq r \leq R \\
\psi_1(r) \equiv \theta_1 - \left( \frac{\sqrt{\theta_1}}{\theta_2-R} (r-\theta_2)^2 \right)& \mbox{if } R \leq r \leq r_1 \\
\psi_2(r) \equiv 1 + \frac{1}{(\theta_3(r-r_2))^2}&\mbox{if } 0 \geq r_1 \end{cases}.
\end{align}
\end{linenomath*}

Using this parametrization, $\theta_1$ gives the peak height of the interaction function, $\theta_2$ gives the location of the peak, and $\theta_3$ controls the rate at which the function descends after the peak. The values $r_1$ and $r_2$ in Equation \eqref{eq:2.7} are the unique real numbers that make $\psi(r)$ and $\frac{d}{dr} \psi(r)$ continuous, given by the solution to the differential equations
\begin{linenomath*}
\begin{align*}
\begin{cases}
\psi_1(r_1) = \psi_2(r_1)\\
\frac{d\psi_1}{dr}(r_1) = \frac{d\psi_2}{dr}(r_1)
\end{cases}.
\end{align*}
\end{linenomath*}
\begin{figure}
\centering
\caption{Behavior of Attraction-Repulsion Interaction Function}
\label{fig:joshInteract}
\includegraphics[width=120mm]{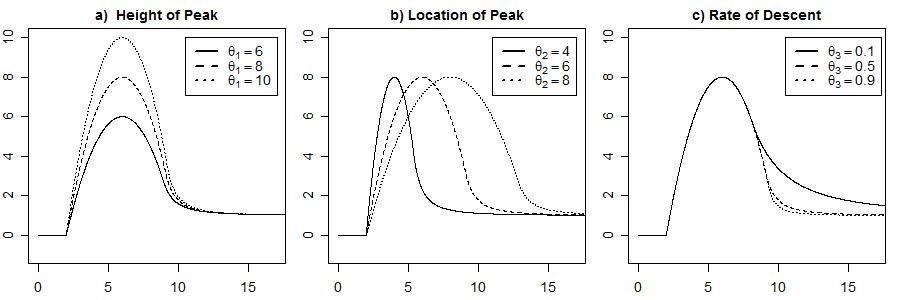}\\
\begin{flushleft}
Examples of the attraction-repulsion interaction function from \citeasnoun**{Goldstein}. \\
a) Demonstrates the effect of changing the peak height parameter $\theta_1$. \\
b) Demonstrates the effect of changing the peak location parameter $\theta_2$. \\
c) Demonstrates the effect of changing the rate of descent parameter $\theta_3$.
\end{flushleft}
\end{figure}
See \citeasnoun**{Goldstein} for details. Examples of the interaction functions under different parameter settings are given in Figure \ref{fig:joshInteract}.

\subsection{Comparison with Existing Approaches}

There have been several other models proposed to account for interaction behavior in animal movement. Let $\boldsymbol{A}_{t_{i}}$ represent the true locations of each of the animals in the group at time $t_i$. \citeasnoun{gautrais2008} utilize a model where the locations at the next time point  $\boldsymbol{A}_{t_{i+1}}$ are only dependent on neighbor's locations at the current time point, so the interaction is a function of $\boldsymbol{A}_{t_{i}}$. Since animals generally interact continuously over time we prefer a model that allows modeling of group behavior based on the joint distribution of the next location of the individuals in the group, resulting in an interaction which is a function of $\boldsymbol{A}_{t_{i+1}}$. This results in a reasonable model even if there are long time lags between the observations. Additionally, we consider direct estimation of model parameters, whereas \citeasnoun{gautrais2008} utilize extensive simulations under different parametrizations followed by analysis of group summary statistics. \citeasnoun{mann2011} discuss Bayesian parameter estimation of a SPP system model, where the interaction term in the model is again assumed to depend only on the system state at the previous time point. However, the analysis is only accurate if the rate of observations matches the rate at which animals update their velocity, implicitly assuming that individuals update their velocities at discrete time points \cite**{mcclintock2014discrete}.

\citeasnoun**{potts2014} propose a similar weighted distribution framework that combines three different aspects of movement, individual movement, the effect of the environment and the interaction with previous behavior of the rest group, to model an individuals next location. These factors are modeled by assuming separability and taking the product of three different parts, the movement process, the environmental desirability weighting function, and the collective interaction which can include all information on group movement up to and including the recent time point. We extend this framework by using ideas from point process statistics to jointly model the probability of members of a group moving to a new location rather than considering the group's recent history. Another recent approach due to \citeasnoun**{Langrock} assumes that animals move around a latent centroid to account for group dynamics in animal movement. The movement of the individuals is modeled as a hidden Markov model with behavioral states. In one state, the animals may be attracted to the group centroid, and follow a biased correlated random walk, whereas in an exploratory state, the individual might follow a correlated random walk. Instead of the latent centroid approach of \citeasnoun**{Langrock}, our method deals with the group dynamics by looking at the pairwise behavior between the individuals directly, allowing for different types of behavior, such as pairs of animals moving together separate from the group. \citeasnoun{mann2011} finds that parameter estimates can be biased if the time lag for the observations does not match the rate at which individuals update their velocities when only the previous locations are considered. Our approach does not have this weakness since we model interaction behavior dependent on the current joint locations of the group of individuals, rather than just the previous locations using point process interaction functions. \citeasnoun**{Johnson_2013} use spatio-temporal point process models to study resource selection, but they do not consider animal interactions.

Our weighted distribution approach provides a general approach to modelling movement interactions that is not affected by the timescale of the observations due to the joint modeling of the locations. This is an improvement over existing methods which model interactions based on the most recent locations under a Markovian assumption. In the case of the guppies, we are able to model individual movement using existing dynamic models, and interaction using existing point process models which provide a natural way of modeling the interaction among points in a plane. Both of these types of models have a large literature basis and this makes modeling accessible.

\section{Model Inference}\label{sec2}

    Next, we describe a Metropolis-Hastings algorithm to perform Bayesian inference. We select priors for each of the parameters that reflect our limited prior information about the model parameters. We will use the same priors for both our simulation study and data analysis. For $\gamma_1$ and $\gamma_2$ we specify conjugate normal priors with zero mean and variance equal to $10^4$, $\pi(\gamma_1)\sim N(0, 10^4)$ and $\pi(\gamma_2)\sim N(0, 10^4)$. For the parameters that are restricted to be positive we specify truncated normal priors, denoted $\textnormal{truncN}(\mu, \sigma^2, B_L)$, with lower bound given by $B_L$ and density proportional to
    \begin{linenomath*}
    \begin{align*}
    f(x|\mu, \sigma^2, B_L) \propto \exp \left( \frac{-(x-\mu)^2}{2\sigma^2} \right) I\{x > B_L\}
    \end{align*}
    \end{linenomath*}
    where $I$ is the indicator function. The priors chosen are given by
     $\beta \sim \textnormal{truncN}(1, 10^4, 0)$, $\sigma^2 \sim \textnormal{truncN}(1, 10^4, 0)$ and $\sigma^2_E \sim \textnormal{truncN}(1, 10^4, 0)$. The parameter R was fixed a priori to be the minimum distance between individuals across all time points, denoted $\hat{R}$. We have additional interaction parameters $\theta_1$, $\theta_2$ and $\theta_3$. For $\theta_1$ and $\theta_2$ we use truncated normal priors; $\theta_1 \sim \textnormal{truncN}(2, 10^4, 1)$ and $\theta_2 \sim \textnormal{truncN}(\hat{R} + 1, 10^4, \hat{R})$. Finally, since the effect of $\theta_3$ on the interaction function is minimal for all $\theta_3$ greater than one (see Figure \ref{fig:joshInteract}) we use a uniform prior on $(0,1)$ for $\theta_3$.

     Inference is straightforward when the point process interactions are not included in the model. For the independent group movement model discussed in Section 2.2, we use variable-at-a-time Metropolis-Hastings. At each iteration of our MCMC algorithm, we first update the unobserved states for each individual at each time point, $\boldsymbol{A_{t_{1:N}}}$, and then each of the model parameters $(\beta, \gamma_1, \gamma_2, \sigma^2, \sigma^2_E)$. The Kalman filter can be used for the model with no interactions but it can not be easily extended to the general case; thus we focus on a more general method for inference.

    We assessed convergence by monitoring Monte Carlo standard errors using the batch means procedures, described in \citeasnoun**{jones2006fixed} and \citeasnoun**{flegal2008markov}, and by comparing kernel density estimates of the posterior of the first half of the chain and the second half of the chain.

    Inference becomes more challenging when interactions are included in the model. Without the interaction function $\psi(\cdot)$, the normalizing constant does not depend on the parameters, so it can be ignored for Bayesian inference. However, the normalizing function in Equation \eqref{eq:density} is a function of all of the model parameters $c(\boldsymbol{\Theta}) = c( \boldsymbol{\Theta_1}, \boldsymbol{\Theta_2} )$. In the Metropolis-Hastings algorithm, using the model likelihood from Equation \eqref{eq:density}, and a proposal density $q(\cdot|\cdot)$ we have acceptance probability:
    \begin{linenomath*}
    \begin{align*}
        \alpha = \text{min}\left( 1, \frac{p(\boldsymbol{\Theta'}) q(\boldsymbol{\Theta'}|\boldsymbol{\Theta}) h\left(\boldsymbol{S_{t_{1:N}}} |\boldsymbol{A_{t_{1:N}}}, \boldsymbol{\Theta'} \right) g\left(\boldsymbol{A_{t_{1:N}}} |\boldsymbol{\Theta'} \right) \psi( \boldsymbol{A_{t_{1:N}}} |\boldsymbol{\Theta'}) c(\boldsymbol{\Theta)}}{p(\boldsymbol{\Theta}) q(\boldsymbol{\Theta}|\boldsymbol{\Theta'}) h\left(\boldsymbol{S_{t_{1:N}}} |\boldsymbol{A_{t_{1:N}}}, \boldsymbol{\Theta} \right) g\left(\boldsymbol{A_{t_{1:N}}} |\boldsymbol{\Theta} \right) \psi( \boldsymbol{A_{t_{1:N}}} |\boldsymbol{\Theta}) c(\boldsymbol{\Theta'})} \right).
    \end{align*}
    \end{linenomath*}
Thus, since the normalizing functions do not cancel out we cannot use Metropolis-Hastings without accounting for them.

    Many methods have been suggested to deal with this issue in the point process literature, however they are often computationally expensive. \citeasnoun{besag1974spatial} proposed an estimation method using psuedo-likelihood which does not work well when there is strong interaction. \citeasnoun{geyer1992constrained} use importance sampling to estimate the normalizing constant, however this method only works if the parameter value used in the importance function is close to the maximum likelihood estimate of the parameter. \citeasnoun**{atchade2013bayesian} propose an MCMC algorithm for Bayesian inference. \citeasnoun{moller2004statistical} gives an overview of several other estimation methods. Here we use the double Metropolis-Hastings (MH) algorithm \cite{Liang}. This is an approximate version of the auxiliary variable M-H algorithm \cite**{moller2006efficient,murray2012} but avoids perfect sampling \cite{propp1996exact} which is not possible from our model. The auxiliary variable is approximately simulated using a nested MH sampler. This avoids estimation of the normalizing constant at the cost of simulating the path using MCMC. The length of the nested MH sampler must be large enough so that the distribution of the auxiliary variable is close to that of a perfect sampler.

    The double MH algorithm \cite{Liang} is
\begin{enumerate}
\item Generate a proposal $\boldsymbol{\Theta'}$ from some proposal distribution $q(\boldsymbol{\Theta} | \boldsymbol{\Theta'})$
\item Generate an auxiliary $\boldsymbol{Y^*} = (\boldsymbol{A^*_{t_{1:N}}},\boldsymbol{S^*_{t_{1:N}}})$ from a kernel with stationary distribution
\begin{linenomath*}
\begin{align*}
\frac{h\left(\boldsymbol{S^*_{t_{1:N}}} |\boldsymbol{A^*_{t_{1:N}}}, \boldsymbol{\Theta'} \right) g\left(\boldsymbol{A^*_{t_{1:N}}} |\boldsymbol{\Theta'} \right) \psi( \boldsymbol{A^*_{t_{1:N}}} |\boldsymbol{\Theta'})}{c(\boldsymbol{\Theta'})}.
\end{align*}
\end{linenomath*}
$\boldsymbol{Y^*}$ is a approximation of a simulated path from the proposal distribution, this is accomplished using a MH algorithm.
\item Accept $\boldsymbol{\Theta'}$ with probability $\alpha=min\left(1, R(\boldsymbol{\Theta}, \boldsymbol{\Theta'})\right)$, where $R(\boldsymbol{\Theta}, \boldsymbol{\Theta'})$ is given by
\begin{linenomath*}
\begin{align*}
 \frac{p\left(\boldsymbol{\Theta'}\right)q\left(\boldsymbol{\Theta'}|\boldsymbol{\Theta}\right)h\left(\boldsymbol{S_{t_{1:N}}} |\boldsymbol{A_{t_{1:N}}}, \boldsymbol{\Theta'} \right) g\left(\boldsymbol{A_{t_{1:N}}} |\boldsymbol{\Theta'} \right) \psi\left( \boldsymbol{A_{t_{1:N}}} |\boldsymbol{\Theta'}\right)} {p\left(\boldsymbol{\Theta}\right)q\left(\boldsymbol{\Theta}|\boldsymbol{\Theta'}\right)h\left(\boldsymbol{S_{t_{1:N}}} |\boldsymbol{A_{t_{1:N}}}, \boldsymbol{\Theta} \right) g\left(\boldsymbol{A_{t_{1:N}}} |\boldsymbol{\Theta} \right) \psi\left( \boldsymbol{A_{t_{1:N}}} |\boldsymbol{\Theta}\right)} H\left(\boldsymbol{\Theta}, \boldsymbol{\Theta'}, \boldsymbol{A^*_{t_{1:N}}},   \boldsymbol{S^*_{t_{1:N}}} \right)
\end{align*}
\end{linenomath*}
and $H\left(\boldsymbol{\Theta}, \boldsymbol{\Theta'}, \boldsymbol{A^*_{t_{1:N}}},   \boldsymbol{S^*_{t_{1:N}}} \right)$ is the ratio
\begin{linenomath*}
\begin{align*}
H\left(\boldsymbol{\Theta}, \boldsymbol{\Theta'}, \boldsymbol{A^*_{t_{1:N}}},   \boldsymbol{S^*_{t_{1:N}}} \right) = \frac{h\left(\boldsymbol{S^*_{t_{1:N}}} |\boldsymbol{A^*_{t_{1:N}}}, \boldsymbol{\Theta} \right) g\left(\boldsymbol{A^*_{t_{1:N}}} |\boldsymbol{\Theta} \right) \psi\left( \boldsymbol{A^*_{t_{1:N}}} |\boldsymbol{\Theta}\right)}{h\left(\boldsymbol{S^*_{t_{1:N}}} |\boldsymbol{A^*_{t_{1:N}}}, \boldsymbol{\Theta'} \right) g\left(\boldsymbol{A^*_{t_{1:N}}} |\boldsymbol{\Theta'} \right) \psi\left( \boldsymbol{A^*_{t_{1:N}}} |\boldsymbol{\Theta'}\right)}.
\end{align*}
\end{linenomath*}
\end{enumerate}

In our model, since none of the parameters can be easily separated from the integration over the unobserved states; the normalizing function is a function of all model parameters. Thus, we need to use the double Metropolis-Hastings algorithm for each parameter update. Therefore, for each parameter update, we use an MH algorithm  to simulate a realization of the unobserved states $\boldsymbol{A^*_{t_{1:N}}}$ and observations $\boldsymbol{S^*_{t_{1:N}}}$ from our model with the proposal parameters, and use this simulation $\boldsymbol{Y^*} = (\boldsymbol{A^*_{t_{1:N}}},\boldsymbol{S^*_{t_{1:N}}})$ to estimate the ratio $H\left(\boldsymbol{\Theta}, \boldsymbol{\Theta'}, \boldsymbol{A^*_{t_{1:N}}},   \boldsymbol{S^*_{t_{1:N}}} \right)$. This requires a simulation of an entire sample path for each new proposal parameter. Note that this estimate is only accurate if the value of $\boldsymbol{\Theta}$ is similar to the value of $\boldsymbol{\Theta'}$, so we elect to use variable at a time updates for all parameters, as opposed to block updates of $\boldsymbol{\Theta}$.

Now we consider the DPPI model from Section 2.4 with the attraction-repulsion interaction function from \citeasnoun{Goldstein}. In each iteration of our double Metropolis-Hastings algorithm we first update the unobserved states, $\boldsymbol{A_{t_{1:N}}}$, using a four-dimensional block Metropolis-Hastings update, where the unobserved state of each fish $j$ at each time point $t_i$, $ \boldsymbol{\alpha}^{(j)}_{t_i}$ consisting of the true x and y locations and instantaneous velocities, is updated one at a time. Next, we update each each parameter ($\beta$, $\gamma_1$, $\gamma_2$, $\sigma^2$, $\sigma^2_E$, $\theta_1$, $\theta_2$, $\theta_3$) one at a time using a double Metropolis-Hastings update. For each parameter, we use a nested MH sampler to generate an auxiliary variable $\boldsymbol{Y^*}$ from the DPPI model using the current parameters in the MCMC chain and the proposed parameter to be updated. For parameters ($\beta$, $\gamma_1$, $\gamma_2$, $\sigma^2$, $\theta_1$, $\theta_2$, $\theta_3$) the auxiliary variable is a simulated realization of the unobserved states $\boldsymbol{Y^*}=\boldsymbol{A^*_{t_{1:N}}}$ and for $\sigma^2_E$ the auxiliary variable also requires a simulated realization of the observations $\boldsymbol{Y^*} = (\boldsymbol{A^*_{t_{1:N}}},\boldsymbol{S^*_{t_{1:N}}})$. Both of these auxiliary variables are generated using a Metropolis-Hastings algorithm.

The length of the nested MH sampler used to generate the auxiliary variable was determined by examining the distances between the simulated realizations of the observed locations $\boldsymbol{S^*_{t_{1:N}}}$ as the length is increased. The length was doubled until the average distance between locations stabilized, resulting in a nested MH sampler of length 200. The double Metropolis-Hastings step is time consuming, since it requires a nested Metropolis-Hastings sampler for each parameter at each MCMC step. Convergence was determined using the same methods as for the independent movement algorithm.

\section{Application to Simulated Data}\label{sec3}

\begin{figure}
\centering
\caption{Simulated Data under Different Settings}
\label{fig:simulation}
\centering
\includegraphics[width=120mm]{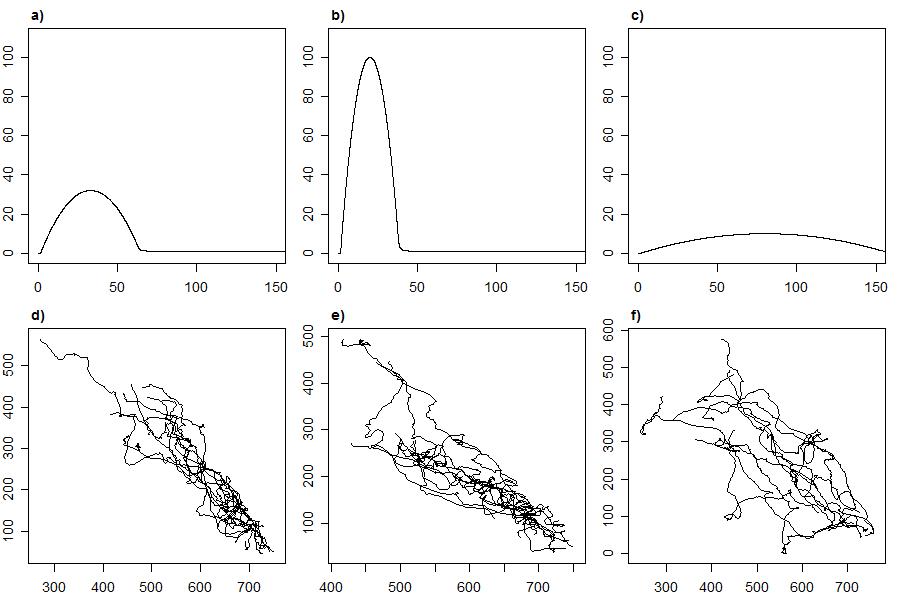}\\
\begin{flushleft}
The attraction-repulsion point process interaction function for the (a)medium, (b)strong, and (c)weak simulated realizations of the model; and plots of the simulated paths for the (d)medium, (e)strong, and (f)weak interactions.
\end{flushleft}
\end{figure}

To test the performance of our double Metropolis-Hastings algorithm we generated simulated paths from our DPPI model and recovered the true parameters. We simulated three group movement paths with starting locations taken from the starting locations of the ten guppies in Figure \ref{fig:SimulatedPaths}(a). In all cases the CTCRW movement parameters $\boldsymbol{\Theta_1}$ were set to the means of the posterior distributions from Section 5 $(\beta=0.15 , \gamma_1=-1.2 ,\gamma_2=1.5 ,\sigma^2=1.7 , \sigma^2_E=0.4)$. The interaction parameters $\boldsymbol{\Theta_2}$ were chosen for three different scenarios. In scenario 1 (medium interaction), we used the posterior mean parameter values from Section 5 $(\theta_1^{(1)}=32, \theta_2^{(1)}=33, \theta_3^{(1)}=0.3)$  to mimic the guppy movement. The parameters in scenario 2 were specified to encourage stronger interaction $(\theta_1^{(2)}=100, \theta_2^{(2)}=20, \theta_3^{(2)}=0.5)$. The parameters in scenario 3 were specified to represent a weaker interaction $(\theta_1^{(3)}=10, \theta_2^{(3)}=80, \theta_3^{(3)}=0.5)$. The interaction functions and simulated paths are plotted in Figure \ref{fig:simulation}. The heights of the interaction functions show that the second set of parameters (Figure \ref{fig:simulation}(b)) results in the strongest interaction, and the third set of parameters (Figure \ref{fig:simulation}(c)) results in the weakest interaction. In the simulated movement paths, it is apparant that Figure \ref{fig:simulation}(c) has less interaction, but it is difficult to compare the strength of attraction between Figures \ref{fig:simulation}(a) and (b) from the plots of the movement paths alone.

We first estimated the parameters using the independent model that assumes that the fish moved independently, as in Section 2.2. The resulting parameter estimates and 95\% equi-tailed credible intervals are given in Table \ref{table:1}.
\begin{table}
\begin{center}
  \caption{Simulated Model Assuming Independent Movement}\label{table:1}
  \begin{tabular}{| l | c  c  c  c  c |}
    \hline
    Interaction Strength & $\beta=0.15$ & $\gamma_1=-1.2$ & $\gamma_2=1.5$ & $\sigma^2=1.7$ & $\sigma^2_E=0.4$ \\ \hline
    Medium &  $.184$ & $-1.25$ & $1.64$ & $1.94$ & $.389$ \\
      &  $(.15, .21)$ & $(-1.57,-0.91)$ & $(1.33, 1.98)$ & $(1.76,2.12)$ & $(.36,.41)$ \\ \hline
    Strong &  $.210 $ & $-1.11 $ & $1.30 $ & $1.92 $ & $.385 $ \\
      &  $(.18, .23)$ & $(-1.40,-0.83)$ & $(1.02, 1.59)$ & $(1.72,2.11)$ & $(.35,.41)$ \\ \hline
    Weak &  $.146 $ & $-1.40 $ & $1.53 $ & $1.75 $ & $.392 $ \\
      &  $(.12, .16)$ & $(-1.78,-1.00)$ & $ (1.13, 1.93)$ & $(1.61,1.93)$ & $(.36,.42)$ \\ \hline
  \end{tabular}
  \begin{flushleft}
  Posterior means and 95\% equi-tailed credible intervals estimated using a variable at a time Metropolis-Hastings algorithm assuming there is no interaction between individuals on the data simulated from a DPPI model with medium $(\theta_1^{(1)}=32, \theta_2^{(1)}=33, \theta_3^{(1)}=0.3)$, strong $(\theta_1^{(2)}=100, \theta_2^{(2)}=20, \theta_3^{(2)}=0.5)$, and weak $(\theta_1^{(3)}=10, \theta_2^{(3)}=80, \theta_3^{(3)}=0.5)$ interaction settings.
  \end{flushleft}
\end{center}
\end{table}
Our credible intervals for for $\gamma_1$, $\gamma_2$, and $\sigma^2_E$ include the true parameters for all of the simulations. However, in the medium and strong attraction scenarios the credible intervals for $\beta$ and $\sigma^2$ do not contain the truth. This indicates that assuming independence when there is actually interaction among the animals can result in biased parameter estimates.

 Next, we used the correct DPPI model to analyze the simulated data. The results are given in Table \ref{table:2}.
\begin{table}
\begin{center}
  \caption{Simulated Model Including Interactions}\label{table:2}
  \begin{tabular}{| l | c  c  c  c  c |}
    \hline
    Interaction Strength & $\beta=0.15$ & $\gamma_1=-1.2$ & $\gamma_2=1.5$ & $\sigma^2=1.7$ & $\sigma^2_E=0.4$ \\ \hline
    Medium &  $.161 $ & $-1.25 $ & $1.64 $ & $1.72 $ & $.404 $ \\
     &  $ (.13, .18)$ & $ (-1.58,-0.90)$ & $(1.30, 2.01)$ & $(1.57,1.86)$ & $ (.37,.43)$ \\ \hline
    Strong &  $.161 $ & $-1.11 $ & $1.32 $ & $1.51 $ & $.413 $ \\
     &  $(.13, .18)$ & $(-1.45,-0.79)$ & $ (1.00, 1.64)$ & $(1.38,1.68)$ & $(.38,.44)$ \\ \hline
    Weak &  $.144 $ & $-1.40 $ & $1.53 $ & $1.74 $ & $.391 $ \\
     &  $(.12, .16)$ & $(-1.82,-1.01)$ & $(1.12, 1.92)$ & $ (1.59,1.91)$ & $ (.36,.42)$ \\ \hline
  \end{tabular}
  \begin{tabular}{| l | c  c  r |}
    \hline
     Interaction Strength& $\theta_1 = (32, 100, 10)$ & $\theta_2 = (33,20,80)$ & $\theta_3 = (0.3,0.3,0.5)$\\ \hline
    Medium &  $37.5 $ & $33.7 $ & $.408 $ \\
      &  $(18.1, 74.4)$ & $ (29.6,39.1)$ & $ (.050, .954)$ \\ \hline
    Strong &  $66.9 $ & $19.4 $ & $.614 $ \\
      &  $(31.0, 134.2)$ & $ (16.6,21.5)$ & $ (.073, .983)$ \\ \hline
    Weak &  $12.4 $ & $78.7 $ & $.359 $ \\
      &  $ (4.0, 33.3)$ & $ (20.2,114.4)$ & $ (.011, .947)$ \\ \hline
  \end{tabular}
  \begin{flushleft}
  Posterior means and 95\% equi-tailed credible intervals estimated using the double Metropolis-Hastings algorithm on the data simulated from a DPPI model with medium, strong, and weak interaction settings.
  \end{flushleft}
  \end{center}
\end{table}
From Table \ref{table:2}, we can see that our algorithm accurately recovers the movement parameters $\boldsymbol{\Theta_1}$ with the exception of $\sigma^2$ which falls just outside the 95\% credible interval in the strong attraction scenario. In Table \ref{table:2}, we are also successful in recovering $\theta_1$ and $\theta_2$, but there is greater uncertainty in these parameter estimates than in the movement parameters. Although the simulated paths looked similar in Figure \ref{fig:simulation}, we are able to distinguish between the medium attraction and the strong attraction scenarios. However the width of the credible interval increases as attraction increases, indicating it is harder to differentiate between levels of attraction as the peak of our attraction-repulsion interaction function increases. For $\theta_3$, the posterior is very similar to the prior distribution, a uniform distribution on $(0,1)$, which indicates that there is not enough information in the simulated data to infer the parameter. To test the effect that having an incorrect estimate for $\theta_3$ would have on the other parameter estimates, the double Metropolis-Hastings algorithm was rerun fixing $\theta_3$ at several different values $\left( \theta_3=0.05 ,\theta_3=0.5 ,\theta_3=0.9 \right)$. The resulting posterior distributions for the other parameters remained consistent with our previous results, so the lack of identifiability of $\theta_3$ does not invalidate our estimates for the other parameters.

\section{Guppy Data}\label{sec4}
We now use our approach to analyze the guppy shoal data of \citeasnoun{Bode2012}, available online \cite**{data}, where the individuals show a tendency to interact, as evident by the shoaling behavior in Figure \ref{fig:SimulatedPaths}(a). Gravel and shade were added in one corner of the tank to attract the guppies, and a group of ten guppies is released in the opposite corner. The full trajectories are observed for the guppies from the time they begin moving towards the destination until the first guppy reaches the target. The guppies were filmed with a standard definition camera, recording 10 frames per second, and tracking software (SwisTrack; \citeasnoun**{lochmatter2008swistrack})was used to obtain the coordinates. One realization of the experiment is plotted in Figure \ref{fig:SimulatedPaths}(a). The experiment was repeated several times, but we focus our analysis on a single realization of the experiment. \citeasnoun{Bode2012} calculated a summary statistic based on angles of direction to estimate the social interactions of a group. A permutation test, which randomly assigned group membership of guppies to artificial experimental trials, found that the social interaction summary statistic was larger in actual groups than in artificially permutated groups in all but 75 out of 10,000 permutations. \citeasnoun{Bode2012} concluded that the guppies do interact socially. Using our approach, we are able to extend the results of \citeasnoun{Bode2012} and directly infer parameter values that reflect this interaction between fish.

We first performed inference using the independent movement model from Section 2.2. Next we used our double Metropolis-Hastings algorithm to estimate the parameters for the DPPI model described in Section 2.3. The priors in both scenarios were selected to be the same as in the simulation example, described in Section 3. The results are presented in Table \ref{table:4}.
 \begin{table}
 \begin{center}
  \caption{Posterior Summary for the Guppy Data}\label{table:4}
  \begin{tabular}{| l | c c c c c |}
    \hline
    Model & $\beta$ & $\gamma_1$ & $\gamma_2$ & $\sigma^2$ & $\sigma^2_E$ \\ \hline
    Indep. &  $.159 $ & $-1.18 $ & $1.51 $ & $1.88 $ & $0.384 $ \\
      &  $(.13, .18)$ & $ (-1.56,-0.80)$ & $ (1.14, 1.89)$ & $ (1.71,2.04)$ & $(0.35,0.41)$ \\ \hline
    Interact &  $.145 $ & $-1.17 $ & $1.51 $ & $1.75 $ & $0.395 $ \\
      &  $(.12, .16)$ & $(-1.58,-0.77)$ & $ (1.12, 1.89)$ & $ (1.60,1.95)$ & $ (0.36,0.42)$ \\ \hline
  \end{tabular}
  \begin{tabular}{| l | c c r |}
    \hline
    Model & $\theta_1$ & $\theta_2$ & $\theta_3$\\ \hline
    Interact &  $32.0 $ & $32.9 $ & $0.304 $ \\
      &  $(15.1, 58.2)$ & $ (23.4, 44.4)$ & $(0.019, 0.921)$ \\ \hline
  \end{tabular}
    \begin{flushleft}
  Posterior means and 95\% equi-tailed credible intervals for the guppy data of \citeasnoun**{Bode2012} assuming no interaction and attraction-repulsion point process interactions, estimated using variable at a time Metropolis-Hastings and the double Metropolis-Hastings algorithm respectively.
  \end{flushleft}
\end{center}
\end{table}
The means of the posterior distributions for the the parameters $\gamma_1$, $\gamma_2$, and $\sigma^2_E$ are almost identical for the independent and the interaction models. However, the estimates for $\beta$ and $\sigma^2$ differ slightly. Our results from the simulation study imply that the independent model estimates could be inaccurate, since the fish interact with each other socially \cite{Bode2012}.  The results for the movement model parameters $\boldsymbol{\Theta_1}$ indicate that there is autocorrelation in the observations over time, the fish tend to move toward the shelter in the upper left corner, and there is appreciable measurement error but it is very small in magnitude, since 0.4 pixels is approximately 0.08 cm. This seems reasonable since the tracking software used by \citeasnoun{Bode2012} is highly accurate \cite**{lochmatter2008swistrack}.

\begin{figure}
\centering
\caption{Pairwise Distance Envelope}
\label{fig:comparison}
\includegraphics[width=110mm]{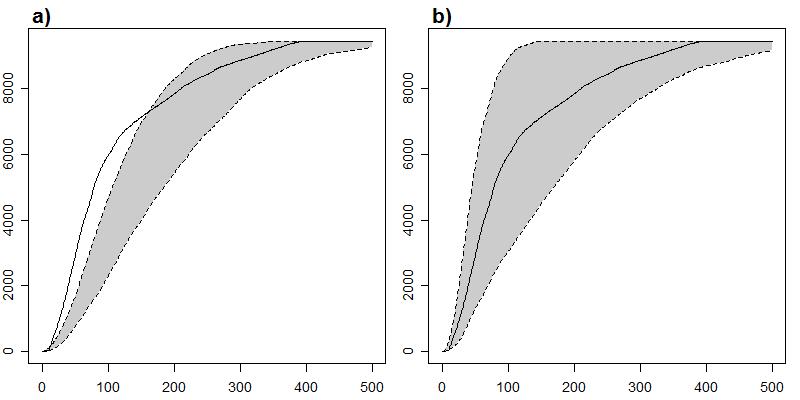}\\
\begin{flushleft}
Estimates of the $K^*(d)$ function for the data compared to 95\% equi-tailed confidence intervals calculated from simulated paths using parameters drawn from the posterior distributions of (a)the CTCRW model assuming no interactions; and (b)The DPPI model with the attraction-repulsion interaction function.
\end{flushleft}
\end{figure}

To compare the independent model and the DPPI model, we analyze the distribution of pairwise distances from simulated realizations of the two models. In point process statistics, Ripley's K function, which is described in \citeasnoun{moller2004statistical}, can be used to analyze the attraction or repulsion between points. The K function, however, requires an estimate for the intensity of the point process, which does not exist in our model since each point has a unique distribution. Instead, we consider the number of pairs of points that lie within a distance of $d$ of each other, a monotone function which starts at 0 and ends at the total number of pairs of points in the process, defined by
\begin{linenomath*}
\begin{align*}
K^*(d)= \sum_{i=1}^N \sum_{k=2}^{K} \sum_{j<k} I\{ \delta (\boldsymbol{\alpha}_{t_{i}}^{(j)}, \boldsymbol{\alpha}_{t_{i}}^{(k)}) <d \}
\end{align*}
\end{linenomath*}
where $I$ represents the indicator function. Larger values of the function indicate that there are more pairs of points within that distance of each other; for example larger values of $K^*(d)$ for small values of d indicate that there is more attraction between points at small scales. To test if our fitted model is capturing the interaction between guppies, we simulate 100 movement paths using draws from the posterior densities of the parameters from the independent movement model and from the DPPI model. We calculate $K^*(d)$ for each of the simulated paths, and create 95\% pointwise envelopes for the K-functions in the two simulation settings by taking the $2.5\%$ and $97.5\%$ quantiles. The $K^*(d)$ function is then calculated for the data and is compared to the envelopes. The result is plotted in Figure \ref{fig:comparison}. The $K^*(d)$ function for the guppy data is above the envelope for the independent movement model at small distances, indicating that there is more attraction between individuals that can be captured in the independent group movement model. When we use the fitted DPPI model with an attraction-repulsion interaction function, the envelope includes the $K^*(d)$ function for the guppy data at all distances, indicating that the inclusion of the interaction function improves the performance of the model in the case of the guppies.

\section{Discussion}\label{sec5}

The movement model with point process interactions we have developed allows us to study group movement of individuals by considering location-based interactions directly. Our double Metropolis-Hastings algorithm for Bayesian inference allows us to accurately estimate parameters. We analyze the movement tracks of a shoal of guppies, which was previously studied using permutation tests and summary statistics in \citeasnoun{Bode2012}, and find that the DPPI model captures the observed pairwise interactions between guppies. We are able to generate paths with similar distributions of pointwise distances between individuals using our model, and show that an independent model fails to do so. We have shown that ignoring interactions of the guppies from \cite{Bode2012} leads to unrealistic group movement paths and inaccuracies in parameter estimates.

 One drawback of our model is that the simulated paths appear less smooth than the actual paths in the data. This could be due to the time-varying behavior of the guppies, which is apparent in Figure \ref{fig:SimulatedPaths}(a), as the guppies change direction during their movement. Further, the guppies do not all start to move at the same time. Some guppies linger at their start location after they have been released. Thus, our assumption of a constant drift shared by all fish may not hold, and including a time-varying drift term that varies across individuals in our CTCRW model might better capture the observed movement behavior. However, this increased flexibility would exacerbate the computational cost, and without incorporating these improvements we are still able to capture the social interactions.

In future work, we would like to consider the impact of unobserved animals interacting with the group. This could potentially result in biased parameter estimates. For example, the strength of the attraction to an individual may be overestimated if there are some unobserved animals moving in a group, or the range of attraction may be overestimated if there are additional unobserved animals between the group members. The locations of unobserved animals could be imputed but this would result in additional computational difficulties, particularly if the number of unobserved individuals is unknown.

 Analysis on group movement mechanics have focused on three main features: collision avoidance at small scales, alignment at medium scales, and attraction at larger scales \cite{gautrais2008}. Our model as presented in Section 2.3 does not explicitly account for the alignment behavior. One method to account for the alignment is to model correlation between the velocities of different individuals as a function of their pairwise distance at the previous time step. \citeasnoun**{katz2011inferring}, however, find that the alignment is automatically induced by the attraction and repulsion behavior, indicating that this might not be necessary to add to the model.

Animal movement models can vary greatly depending on the species being considered. In this case, we have only analyzed the movement of guppies, so the results of our analysis may not extend directly to other animals with different types of interactions. The flexibility to choose a dynamic movement and interaction function provides the potential to model a variety of methods of movement, especially when there is prior knowledge of the animal's behavior.

\section*{Acknowledgements}
\thispagestyle{empty}

We would like to acknowledge the insightful and constructive comments provided by two anonymous reviewers and the associate editor which have clarified and improved the manuscript. This material is based upon work supported by the National Science Foundation under Grant No. 1414296 (Russell and Hanks).

\bibliography{extracted2}

\end{document}